\documentclass{article}
\usepackage{graphicx} 
\usepackage{siunitx}
\usepackage{listings}
\usepackage{hyperref}
\usepackage{nomencl} 
\makenomenclature
\usepackage{amsmath,amssymb}
\usepackage{algorithm}
\usepackage{algpseudocode}
\usepackage{bm}
\usepackage{booktabs}
\usepackage[letterpaper, total={6in, 9in}]{geometry}
\usepackage{breqn}
\providecommand{\keywords}[1]{
  \small
  \textbf{\hfill{Key words}:---} #1
}

\usepackage{pgfplots}
\usepackage{orcidlink}

\pgfplotsset{compat=1.18}

\title{Reduced Precision Diffusion Synthetic Acceleration for S$_N$ Neutron Transport in LLNL's ARDRA using Hypre\footnote{Approved for unlimited release: LLNL-CONF-2024048}}

\newcommand{\lp}{^{(l+1)}}
\newcommand{\la}{^{(l)}}
\newcommand{\lph}{^{(l+1/2)}}

\author{Joanna Piper Morgan\orcidlink{0000-0003-1379-5431},
Mario I. Ortega\orcidlink{0000-0002-5498-9577}}

\date{
    \small{
    Lawrence Livermore National Laboratory, Livermore, CA 94551\\
    \{morgan83, ortega31\}@llnl.gov
    }
}

\begin{document}

\maketitle

\begin{abstract}
Fast solution times in production discrete ordinates transport codes often require the use of diffusion synthetic acceleration (DSA).
A good implementation of DSA will substantially reduce the number of iterations to converge the transport solution at the cost of solving a diffusion-like linear system at each iteration.
Often fewer transport iterations will lead to smaller times to solution.
Recent work has shown the promise of reduced precision preconditioning.
Time and memory cost can be decreased as long as the reduced precision does not substantially degrade the convergence behavior of the higher precision linear system being solved.
In this paper, we discuss the algorithmic implications of accelerating transport iterations in double precision (64 bit) while using a reduced precision (32 bit) DSA solve.
We initially implement a stand alone code to investigate any theoretical limitations of reduced precision DSA in slabs.
Then we implement reduced precision DSA in ARDRA, Lawrence Livermore National Laboratory's (LLNL) discrete ordinance neutral particle transport code, coupled with {\it hypre}, LLNL's scalable linear solver and multigrid methods library capable of mixed precision linear solves
Then we solve various neutron transport problems of interest.
We find that in most circumstances, DSA acceleration in reduced precision has little to no impact on the acceleration properties of DSA when used with source iteration in most circumstances.
However, as tighter tolerances are required, reduced precision DSA may fail to successfully accelerate the transport solve and increase time to solution.

\end{abstract}


\section{Introduction}\label{sec:1}

Deterministic neutron transport methods approximate the neutron angular flux in a six-dimensional phase space (three spatial variables, two angular variables, and one in energy) as a function of time.
For production discrete ordinates (S$_N$) codes, memory usage is a persistent concern for large high fidelity calculations.
As the number of spatial zones, angular quadrature points, or energy groups increase, the number of unknowns in the neutron angular flux increases exponentially.
As computer architectures move toward smaller on-node memory availability, memory pressure is one of the primary constraints on problem fidelity and physics even on massive supercomputers.
Discrete ordinates typically use source iteration (SI) to converge the neutron scattering and fission sources.
By sweeping across the problem domain for all angles and energy groups, these transport sweeps damp high-frequency error modes effectively.
However, it is well known that source iteration alone may converge slowly when scattering is strong since source iteration does not effectively dampen low-frequency error modes.
Effective suppresion of low-frequency modes can be achieved by using some sort of preconditioner. 
In discrete ordinates neutron transport the diffusion synthetic acceleration (DSA) preconditioner is a standard acceleration method for source iteration in diffusive transport regimes \cite{alcouffe1977dsa,adams2002fast}.
While DSA is an effective preconditioner, the solution of the DSA equations can require a large memory footprint and increase computational time in many large ARDRA calculations. Recent work has shown the promise of reduced precision preconditioning for neutron diffusion k-effective problems \cite{CHEREZOV2024110575}.
In this work, we expand the application to solving the DSA equations in low precision to accelerate a full precision transport sweep.
Finding ways to reduce the memory footprint of ARDRA's neutron transport calculations may allow for more physics to be implemented or higher fidelity models considered.

DSA applies a low-order diffusion correction that targets the slowly converging error components with an additive correction.
After a transport sweep produces a new scalar-flux estimate, DSA forms the residual from the previous accelerated scalar flux, applies the low-order DSA operator, $C_{\mathrm{DSA}}$, and adds the resulting correction back to the previous accelerated iterate.
In simplified notation, if $\phi^{(l+1/2)}$ is the post-sweep scalar flux and $\phi^{(l)}$ is the stored previous accelerated flux, Ardra forms a DSA-corrected increment and updates
\begin{equation}
  \phi^{(l+1)} = \phi^{(l)} + C_{\mathrm{DSA}}\left(\phi^{(l+1/2)}-\phi^{(l)}\right).
\end{equation}
Because this correction is lower order than the transport solve, it is a plausible candidate for reduced precision: the transport discretization, cross-section processing, and convergence tests can remain double precision while the DSA matrix, right-hand side, and correction solve use single precision.

Reduced precision is not numerically invisible.
The Institute of Electrical and Electronics Engineers (IEEE) single-precision format has a unit roundoff of approximately $6 \times 10^{-8}$, compared with approximately $1 \times 10^{-16}$ for double precision \cite{ieee754,higham2002accuracy}.
A single-precision DSA solve therefore introduces larger errors in each low-order correction, with their effect governed by coupled transport--DSA conditioning and the inner DSA solver tolerance.
Previous work has exmained the impact of reduced precision in Newtons method \cite{ctkelly_reduced_precision}.
The practical question is whether those errors materially change convergence or quantities of interest in production calculations.
Hereafter, FP32 and FP64 denote 32-bit single-precision and 64-bit double-precision floating-point arithmetic, respectively.

ARDRA interfaces with {\it hypre}\footnote{Pronounced like ``hyper''.}, LLNL's library of high-performance linear solvers and preconditioners \cite{falgout2002hypre}.
{\it hypre} can be configured for single, double, long-double, or mixed precision as of curent released versions.
In mixed-precision builds, {\it hypre} generates precision-specific C interfaces with suffixes such as \texttt{\_flt} and \texttt{\_dbl} and provides a global runtime precision selector.
This paper uses those interfaces to run DSA in single precision while preserving the double precision transport state in ARDRA.

In the remainder of this work we analyze the impact of reduced precision DSA in a one-dimensional stand alone solver, we describe the ARDRA/{\it hypre} implementation, and then we examine the performance of reduced precison DSA in ARDRA on various fixed-source and criticality test problems.

\section{Initial Analysis}

To study a mixed precision DSA solver we implement a stand-alone one-dimensional homogeneous transport model with source iteration and DSA to isolate the algorithmic effect of reduced precision.
The double-precision transport sweep was treated as the reference high-order operation.
Algorithm \ref{alg:reduced_precision} shows the DSA correction process.
The uncorrected double-precision scalar flux was determined using an SI transport sweep. This scalar flux was then used to determine the single precision residual by converting the low-order right-hand side from double to single, applying the single-precision diffusion solve, and converting the correction back to double before updating the transport iterate.
Both the diamond-difference and simple-corner-balance DSA operators are fully consistent with their corresponding transport discretizations through Larsen's four-step construction \cite{adams2002fast}.

\begin{algorithm}
\begin{algorithmic}[1]

    \State initial guess for $\phi$

    \While{not converged}

        \State $\phi_{64}\lph = T \phi\la $ \Comment{SI transport sweep}

        \State $r_{64} = \phi_{64}\lph-\phi_{64}\la$ \Comment{form DSA residual in double}

        \State $r_{32} = R_{64\rightarrow 32} r_{64}$ \Comment{\texttt{static\_cast} from double to float}

        \State $d_{32} = D_{32}r_{32}$ \Comment{DSA solve in float}

        \State $\phi_{64}\lp = \phi_{64}\lph + J_{32\rightarrow 64}d_{32}$ \Comment{update in double}

        \State $l = l + 1$

    \EndWhile
    
    \caption{Reduced precision DSA solve for high precision transport.}
    \label{alg:reduced_precision}
\end{algorithmic} 
\end{algorithm}

Reducing only the precision of the low-order correction does not necessarily impose a single-precision floor on the double-precision transport solution.
Let $B$ denote the transport error operator and $C$ the exact DSA correction operator, so that the FP64 error propagator is $E_{64}=B+C(B-I)$.
Rounding in matrix assembly, data conversion, and the low-order solve replaces $C$ with an approximate operator $\widetilde C_{32}$, giving
\begin{equation}
e^{(l+1)} = \left[B+\widetilde C_{32}(B-I)\right]e^{(l)}
             + \eta^{(l)}_{32},
\end{equation}
where $\eta^{(l)}_{32}$ contains the remaining floating-point error in applying the correction.
Because the DSA right-hand side decreases with the transport residual, this perturbation generally decreases as the iteration converges; consequently, the FP64 fixed point can be retained even though the convergence factor is perturbed.
Stagnation or divergence remains possible if the FP32 solve is inaccurate relative to the residual, if the low-order system is sufficiently ill-conditioned, or if the perturbed propagator has spectral radius greater than or equal to one.

\begin{figure}
    \centering
    \includegraphics[width=\linewidth]{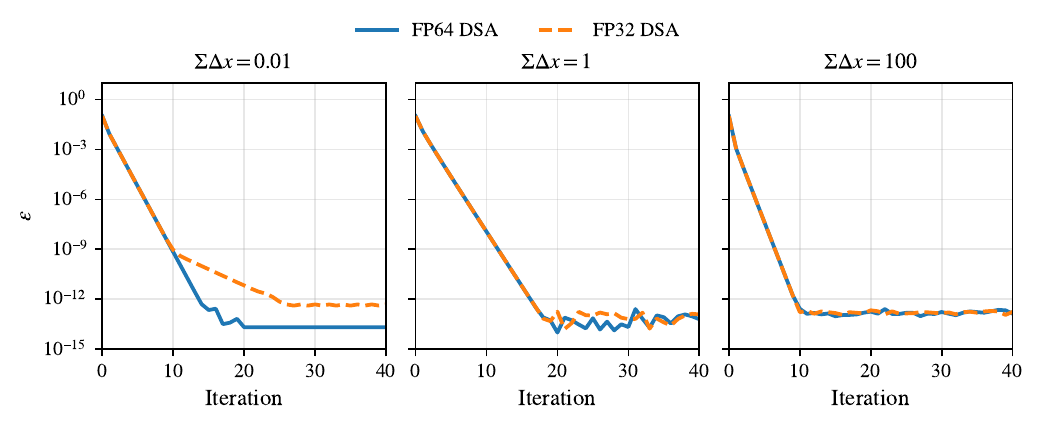}
    \caption{Residual histories for the one-dimensional diamond-difference model. With FP64 residual formation and flux updates, FP32 DSA closely follows the FP64 reference to residuals well below FP32 unit roundoff.}
    \label{fig:simple-residual}
\end{figure}

Figure~\ref{fig:simple-residual} is consistent with this interpretation.
When residual formation and the flux update remain in FP64, the FP32 DSA calculation follows the FP64 convergence history to approximately $10^{-13}$ for all three optical thicknesses; no floor near FP32 unit roundoff is observed.
The DSA system is solved by direct lower--upper (LU) decomposition, so these results isolate finite-precision matrix storage and solution from an iterative solver tolerance.

\begin{figure}
    \centering
    \includegraphics[width=\linewidth]{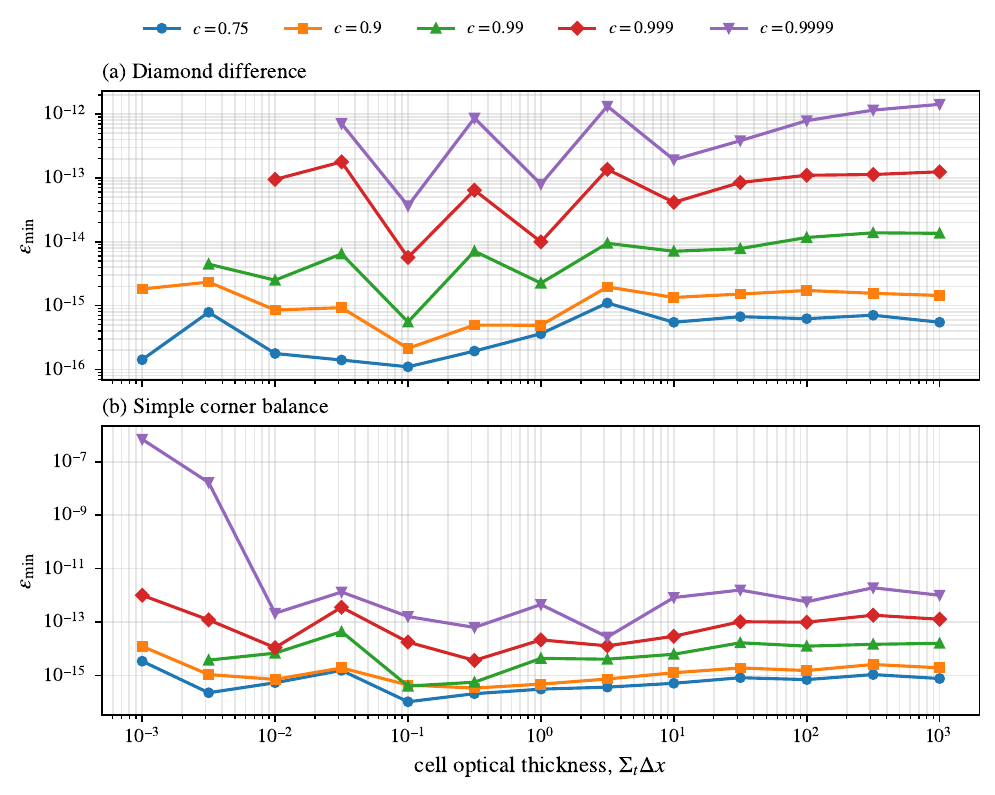}
    \caption{Minimum FP32 DSA residual over the final 20 iterations as a function of cell optical thickness, $\Sigma_t\Delta x$, for several within-group scattering ratios. The top panel uses diamond difference and the bottom panel uses simple corner balance; in both cases the DSA discretization is fully consistent with the transport discretization.}
    \label{fig:single-group-optical-thickness}
\end{figure}

To separate the effect of the coupled transport--DSA problem from a single hand-picked material, the single-group model was also run over a grid of within-group scattering ratios and cell optical thicknesses.
For each case, the total cross section was chosen from $\Sigma_t \Delta x$ and the removal cross section was then set by $\Sigma_a=(1-c)\Sigma_t$, where $c=\Sigma_s/\Sigma_t$.
The sweep used $c\in\{0.75,0.9,0.99,0.999,0.9999\}$ and $\Sigma_t\Delta x\in[10^{-3},10^{3}]$.
For each case, the FP64 transport-plus-DSA error propagator was assembled as $E_{T+D}=B+C(B-I)$, where $B$ is the transport error operator and $C$ is the DSA correction operator.
The coupled condition estimate $\kappa_2(I-E_{T+D})$ was retained as a diagnostic, but Fig.~\ref{fig:single-group-optical-thickness} reports the minimum FP32 DSA residual over the final 20 iterations against $\Sigma_t\Delta x$ directly for both diamond difference and simple corner balance.
This view is more useful for the model problem because it separates discretization scale from the scattering ratio.
Most finite cases converge to residuals between $10^{-16}$ and $10^{-12}$. The largest exception is an optically thin, highly scattering simple-corner-balance case with a residual of $6.90\times10^{-7}$.
Six diamond-difference cases and one simple-corner-balance case produce non-finite iterates and are omitted from the logarithmic plot. These exceptions occur where the FP32 DSA matrix is most ill-conditioned, supporting loss of accuracy or stability in the low-order solve rather than a universal precision floor.

\section{Implementation}

ARDRA selects the DSA precision at run time through the \texttt{dsa/hypre\_precision} option, which accepts \texttt{single}, \texttt{double}, or a build-dependent default. Single precision requires \textit{hypre} to be configured with \texttt{HYPRE\_ENABLE\_MIXED\_PRECISION}. 
A scoped guard queries the active precision with \\\texttt{HYPRE\_GetGlobalPrecision}, selects \texttt{HYPRE\_REAL\_SINGLE} or \texttt{HYPRE\_REAL\_DOUBLE} with \texttt{HYPRE\_SetGlobalPrecision}, and restores the previous setting after each DSA operation. 
Matrix and vector creation, assembly, solver setup, solution, and destruction therefore use a consistent precision without changing the precision of the surrounding transport calculation. The existing \textit{hypre} algorithms are unchanged: one-dimensional systems use structured cyclic reduction, while the multidimensional paths retain the configured structured or semi-structured solver, including semicoarsening multigrid (SMG), parallel semicoarsening multigrid (PFMG), preconditioned conjugate gradient (PCG), Jacobi, and Boomer algebraic multigrid (BoomerAMG).

ARDRA continues to form DSA coefficients and right-hand sides in FP64.
At the \textit{hypre} interface, the single-precision path explicitly casts these arrays to temporary \texttt{float} buffers and loads them through the \texttt{HYPRE\_Struct*BoxValues\_flt} or \texttt{HYPRE\_SStruct*BoxValues\_flt} matrix and vector interfaces. The FP64 path calls the matching \texttt{\_dbl} interfaces.
After the solve, \\\texttt{HYPRE\_StructVectorGetBoxValues\_flt} or \texttt{HYPRE\_SStructVectorGetBoxValues\_flt} returns the FP32 correction, which is cast back to \texttt{double} before ARDRA updates the transport flux.
So, reduced precision is confined to storage and operations within the DSA solve. The current FP32 transfer path is implemented only for central processing unit (CPU) builds.

\section{Verification and Test Problems}

Given the performance of reduced-precision DSA in a stand-alone discrete ordinates transport code, we next examine criticality and fixed-source problems solved with ARDRA.
We compare unaccelerated calculations with FP32 and FP64 DSA using the same outer convergence criterion.
All calculations were performed on RZWhippet and RZHound \cite{llnl_2026_hpc}, identical LLNL clusters with two Intel Xeon Platinum 8479H processors (112 cores total) and 256~gigabytes of memory per node.

\subsection{Ardra's Automated Test Suite}

As an implementation-level verification, we ran ARDRA's automated test suite with the transport calculation in FP64 and the \textit{hypre} DSA solve in FP32.
The suite spans one-, two-, and three-dimensional fixed-source, time-dependent, criticality, adjoint, multigroup, mesh-refinement, and Message Passing Interface (MPI) parallel-decomposition problems.
Each deck first executes ARDRA and, if execution succeeds, compares the generated output with the established FP64 baseline.
The relative comparison tolerance was $10^{-5}$.
Of the 282 cases that completed, 243 passed both execution and comparison and 39 failed either stage.
Two additional cases did not complete before the run ended and are reported separately rather than classified as passes or failures.


Eighteen failures occurred during ARDRA execution.
In every case, the calculation was a two-dimensional criticality problem using the structured-grid SMG solver, and the eigenvalue became non-finite on the first outer iteration.
Eight of these cases are variants of the same linear-discontinuous finite-element (LDFE) problem, four are variants of a mesh-coarsening problem, and the remaining six exercise other combinations of multigroup, quadrature, and adaptive-mesh options.
The two incomplete cases provide consistent additional evidence: a criticality calculation produced a non-finite eigenvalue, while a fixed-source calculation continued for more than 13,900 block-Jacobi iterations with a non-finite relative change.
We believe these failures arise from loss of numerical stability in the FP32 SMG correction for a subset of poorly scaled or ill-conditioned two-dimensional DSA systems.
The immediate appearance of non-finite values, and the identical behavior of serial and parallel LDFE variants, argue against gradual accumulation of roundoff or an MPI decomposition error.

The other 21 failures completed the ARDRA calculation but exceeded the differencing tolerance.
They span one-dimensional multigroup, two-dimensional volume-source, three-dimensional fixed-source, criticality, adjoint, mesh-refinement, and parallel-decomposition calculations.
Absolute differences between the FP32 and reference data are from $1.01\times10^{-5}$ to $1.07\times10^{5}$ and relative differences from $2.00\times10^{-5}$ to $1.16\times10^{-1}$.
These are converged FP32-DSA solutions rather than solver failures, but they show that the reduced-precision configuration can measurably perturb sensitive integral responses even when the transport iteration terminates normally.

The larger relative differences are partly a consequence of the scale of the reference quantity.
For a reference value $q_{64}$ and an FP32-DSA result $q_{32}$, the reported relative difference is $|q_{32}-q_{64}|/|q_{64}|$; consequently, a modest absolute perturbation can appear large where a group flux or leakage is small.
For example, the largest relative difference, 11.6\%, results from an absolute leakage difference of $9.45\times10^{-5}$ relative to a reference value of $8.16\times10^{-4}$.
The one-dimensional analysis shows that a well-conditioned FP32 correction can retain the FP64 fixed point, whereas sufficiently ill-conditioned low-order systems lose accuracy or become non-finite.
The suite's large solution changes and non-finite failures are qualitatively consistent with the latter regime, while its large relative errors in small-valued quantities primarily reflect normalization by a small reference value.
These observations do not quantitatively establish a common mechanism; further work with simplified models is warrented to investigate this.

\subsection{Selected Inputs}

We examine four (one fixed point and three criticality) calculations with reduced precision DSA. We find that for standard ARDRA criticality calculations, FP32 DSA does not substantially degrade acceleration of the calculations while sometimes incurring additional time to solution. ARDRA memory edits indicate the total amount of memory required for DSA is reduced by a factor of half. When FP32 DSA degrades acceleration, the number of iterations required to converge the problem increases adding additional solution overhead. We note that all of these calculations are done on CPU machines. It is possible that GPU implementations of ARDRA would see larger benefits from FP32 DSA and this is future work.

\subsubsection{Three-Dimensional National Ignition Facility (NIF) Fixed Source Model}

The National Ignition Facility (NIF) fixed-source model represents a three-dimensional portion of the target chamber and surrounding shielding, including the chamber, shotcrete, beam ports, direct-drive port plugs, polyethylene plugs, utility penetrations, and detector regions.
An isotropic deuterium--tritium (D--T) point source (representing a NIF shot) at the target position emits $10^{7}$ neutrons in the highest-energy group.
The Cartesian domain is $-590 \leq x \leq -118$~cm, $-472 \leq y \leq 0$~cm, and $-472 \leq z \leq 472$~cm.
It is discretized with $200\times200\times400$ (\num{16e6}) cells decomposed over $4\times4\times7=112$ spatial MPI ranks.
We use $S_4$ angular quadrature for the transport iterations, a second-order Legendre expansion of scattering, and Evaluated Nuclear Data Library (ENDL) 2009 version 5.0 data.

The three neutron groups span 14.407--10.120~MeV, 10.120--1.0245~MeV, and 1.0245~MeV--0.13068~eV, respectively.
Diffusion synthetic acceleration (DSA) is applied only to the lowest-energy group.
We use FP32 and FP64 BoomerAMG DSA corrections for the reduced and normal conditions respectively.

\begin{figure}[]
  \centering
  \includegraphics[width=\linewidth]{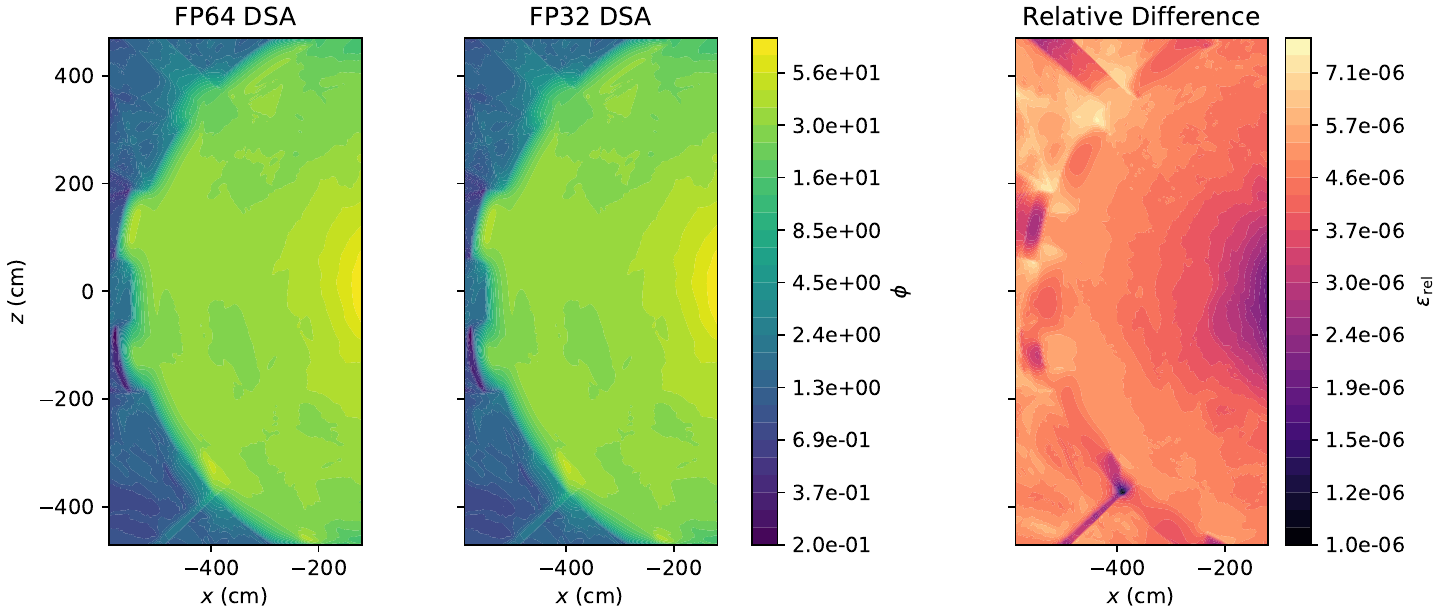}
  \caption{Energy-integrated scalar flux $\phi$ for the FP64- and FP32-DSA steady-state NIF calculations on the $x$--$z$ cell plane at $y=-1.18$~cm, followed by the pointwise relative difference $\epsilon_{\mathrm{rel}}=|\phi_{32}-\phi_{64}|/|\phi_{64}|$. The two flux panels use a common logarithmic scale.}
  \label{fig:nif-steady-flux}
\end{figure}

\begin{figure}[]
  \centering
  \includegraphics[width=\linewidth]{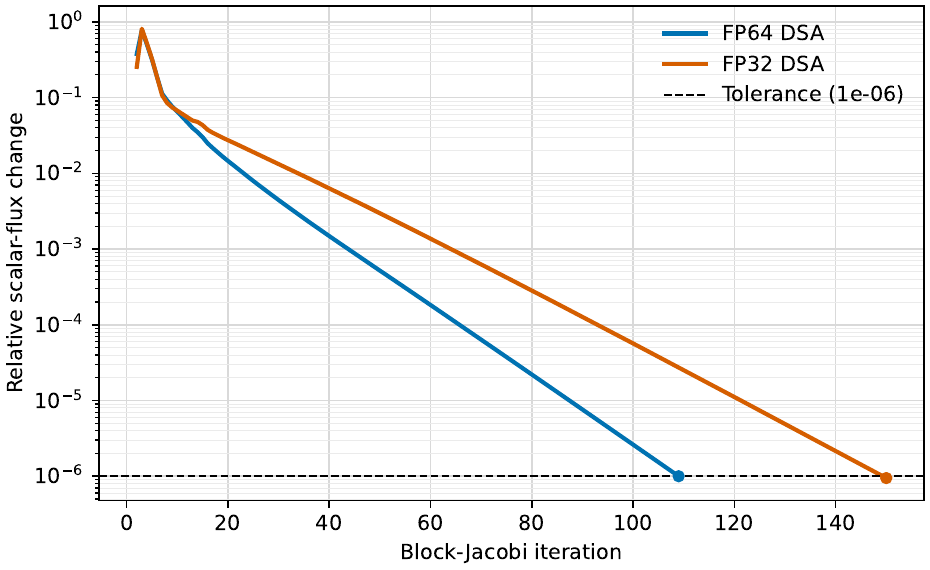}
  \caption{Relative scalar-flux change during block-Jacobi iteration for the steady-state NIF calculation. Each iteration advances the coupled three-group system, with DSA applied to the thermal group. FP64 DSA reaches the $10^{-6}$ tolerance in 109 iterations, whereas FP32 DSA requires 150 iterations.}
  \label{fig:nif-pbj-convergence}
\end{figure}

Both steady-state calculations reached the common $10^{-6}$ outer tolerance, but FP32 DSA required 150 block-Jacobi iterations over groups compared with 109 for FP64 DSA.
The resulting energy-integrated scalar-flux fields remain close: over all \num{16e6} cells, $\lVert\phi_{32}-\phi_{64}\rVert_2/\lVert\phi_{64}\rVert_2=4.56\times10^{-6}$, the largest absolute difference is $2.15\times10^{-4}$, and the largest pointwise relative difference is $8.67\times10^{-6}$.
Figure~\ref{fig:nif-steady-flux} shows the two converged fields and their relative difference on the cell-centered plane nearest the $y=0$ symmetry boundary.
These results show a small change in the converged solution but a substantial, 38\% increase in outer iterations.
The coupled multigroup convergence histories in Fig.~\ref{fig:nif-pbj-convergence} show that the FP32 calculation has a smaller initial relative change but a slower asymptotic decay.
We think that the FP32 degrades the accuracy of the thermal-group DSA correction enough to reduce its effectiveness without changing the converged FP64 transport fixed point appreciably.

\subsubsection{Three-Dimensional Jezebel }
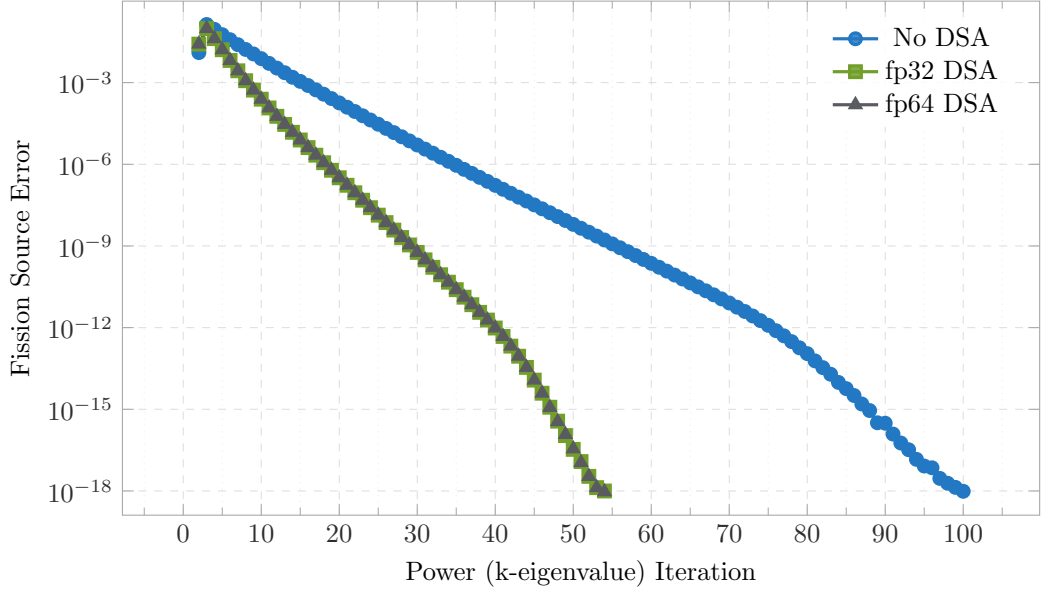
\begin{figure}
    \centering
    \definecolor{llnlblue}{HTML}{2378C3}
\definecolor{llnlgreen}{HTML}{719F2A}
\definecolor{llnlgray}{HTML}{595B60}
\definecolor{llnlorange}{HTML}{D55E00}
\definecolor{llnldarkblue}{HTML}{005DAB}
\definecolor{llnlpurple}{HTML}{CC79A7}
\definecolor{llnllightblue}{HTML}{56B4E9}
\definecolor{llnlblack}{HTML}{000000}

\begin{tikzpicture}
\begin{axis}[
    width=0.90\linewidth,
    height=0.55\linewidth,
    xlabel={Power (k-eigenvalue) Iteration},
    ylabel={Fission Source Error},
    title style={color=llnlblack, font=\bfseries},
    axis line style={draw=gray!70},
    tick style={draw=gray!70},
    tick label style={color=gray!35!black},
    legend pos=north east,
    legend style={draw=none, fill=white, fill opacity=0.85, text opacity=1},
    grid=major,
    major grid style={draw=gray!25, dashed},
    minor grid style={draw=gray!15, dotted},
    minor x tick num=1,
    ymode=log,
    log basis y=10,
    minor y tick num=8,
    grid=both,
    ymin=1.29009430886e-19,
    ymax=0.985038883247
]

\addplot[color=llnlblue, line width=1.2pt, mark=*, mark size=2.2pt]
coordinates {
    (2,0.01266815)
    (3,0.1365032)
    (4,0.09028022)
    (5,0.0576395)
    (6,0.03769797)
    (7,0.02474665)
    (8,0.01653699)
    (9,0.01132044)
    (10,0.007592448)
    (11,0.005094808)
    (12,0.003428533)
    (13,0.002314475)
    (14,0.001573108)
    (15,0.001111124)
    (16,0.0007835944)
    (17,0.0005421949)
    (18,0.0003752366)
    (19,0.0002598689)
    (20,0.0001800969)
    (21,0.0001250241)
    (22,8.697027e-05)
    (23,6.062145e-05)
    (24,4.234063e-05)
    (25,2.96309e-05)
    (26,2.077612e-05)
    (27,1.459439e-05)
    (28,1.027013e-05)
    (29,7.239331e-06)
    (30,5.111108e-06)
    (31,3.613991e-06)
    (32,2.559028e-06)
    (33,1.81443e-06)
    (34,1.288085e-06)
    (35,9.154834e-07)
    (36,6.513636e-07)
    (37,4.639053e-07)
    (38,3.307012e-07)
    (39,2.359466e-07)
    (40,1.684747e-07)
    (41,1.203849e-07)
    (42,8.609814e-08)
    (43,6.190659e-08)
    (44,4.482219e-08)
    (45,3.227212e-08)
    (46,2.319931e-08)
    (47,1.667737e-08)
    (48,1.198966e-08)
    (49,8.620672e-09)
    (50,6.198968e-09)
    (51,4.457804e-09)
    (52,3.206096e-09)
    (53,2.306199e-09)
    (54,1.65886e-09)
    (55,1.193352e-09)
    (56,8.584722e-10)
    (57,6.176923e-10)
    (58,4.443098e-10)
    (59,3.196395e-10)
    (60,2.299299e-10)
    (61,1.653091e-10)
    (62,1.188565e-10)
    (63,8.537356e-11)
    (64,6.133574e-11)
    (65,4.396461e-11)
    (66,3.146217e-11)
    (67,2.249271e-11)
    (68,1.605028e-11)
    (69,1.137646e-11)
    (70,8.049458e-12)
    (71,5.634897e-12)
    (72,3.921188e-12)
    (73,2.682361e-12)
    (74,1.809504e-12)
    (75,1.215998e-12)
    (76,7.712846e-13)
    (77,5.001163e-13)
    (78,3.062054e-13)
    (79,1.81695e-13)
    (80,1.109705e-13)
    (81,6.001898e-14)
    (82,3.39441e-14)
    (83,1.941773e-14)
    (84,9.719083e-15)
    (85,5.753713e-15)
    (86,3.212445e-15)
    (87,1.565697e-15)
    (88,8.989289e-16)
    (89,3.202431e-16)
    (90,3.080755e-16)
    (91,1.230904e-16)
    (92,5.773419e-17)
    (93,3.273399e-17)
    (94,1.453121e-17)
    (95,8.240257e-18)
    (96,7.108005e-18)
    (97,2.897551e-18)
    (98,1.921324e-18)
    (99,1.356863e-18)
    (100,9.732477e-19)
};
\addlegendentry{No DSA}

\addplot[color=llnlgreen, line width=1.2pt, mark=square*, mark size=2.2pt]
coordinates {
    (2,0.02594028)
    (3,0.09682985)
    (4,0.04119781)
    (5,0.01614009)
    (6,0.006524147)
    (7,0.002718891)
    (8,0.001174954)
    (9,0.0005311106)
    (10,0.0002462489)
    (11,0.0001174447)
    (12,5.785122e-05)
    (13,2.9169e-05)
    (14,1.49805e-05)
    (15,7.804627e-06)
    (16,4.099741e-06)
    (17,2.160869e-06)
    (18,1.141175e-06)
    (19,6.036432e-07)
    (20,3.197434e-07)
    (21,1.695854e-07)
    (22,9.004347e-08)
    (23,4.784965e-08)
    (24,2.544435e-08)
    (25,1.353706e-08)
    (26,7.205061e-09)
    (27,3.836057e-09)
    (28,2.042884e-09)
    (29,1.088114e-09)
    (30,5.796527e-10)
    (31,3.088055e-10)
    (32,1.64523e-10)
    (33,8.762387e-11)
    (34,4.664375e-11)
    (35,2.479197e-11)
    (36,1.314025e-11)
    (37,6.954422e-12)
    (38,3.637761e-12)
    (39,1.89657e-12)
    (40,9.682231e-13)
    (41,4.742392e-13)
    (42,2.092679e-13)
    (43,8.913416e-14)
    (44,3.457693e-14)
    (45,1.182162e-14)
    (46,3.891084e-15)
    (47,1.191845e-15)
    (48,3.66766e-16)
    (49,1.082036e-16)
    (50,3.37272e-17)
    (51,1.204575e-17)
    (52,3.441155e-18)
    (53,1.325662e-18)
    (54,9.986233e-19)
};
\addlegendentry{fp32 DSA}

\addplot[color=llnlgray, line width=1.2pt, mark=triangle*, mark size=2.2pt]
coordinates {
    (2,0.02594028)
    (3,0.09682989)
    (4,0.04119772)
    (5,0.01614008)
    (6,0.006524125)
    (7,0.002718886)
    (8,0.001174951)
    (9,0.0005311083)
    (10,0.0002462482)
    (11,0.0001174443)
    (12,5.785099e-05)
    (13,2.916889e-05)
    (14,1.498046e-05)
    (15,7.804597e-06)
    (16,4.099732e-06)
    (17,2.160865e-06)
    (18,1.141172e-06)
    (19,6.036414e-07)
    (20,3.19743e-07)
    (21,1.695852e-07)
    (22,9.004332e-08)
    (23,4.784957e-08)
    (24,2.54443e-08)
    (25,1.353706e-08)
    (26,7.205047e-09)
    (27,3.836052e-09)
    (28,2.042879e-09)
    (29,1.088114e-09)
    (30,5.796539e-10)
    (31,3.088025e-10)
    (32,1.64525e-10)
    (33,8.762334e-11)
    (34,4.664249e-11)
    (35,2.479363e-11)
    (36,1.313831e-11)
    (37,6.956898e-12)
    (38,3.635919e-12)
    (39,1.897284e-12)
    (40,9.684005e-13)
    (41,4.736614e-13)
    (42,2.096107e-13)
    (43,8.910211e-14)
    (44,3.451968e-14)
    (45,1.182842e-14)
    (46,3.910913e-15)
    (47,1.172404e-15)
    (48,3.514978e-16)
    (49,1.185423e-16)
    (50,3.529426e-17)
    (51,1.108288e-17)
    (52,3.425211e-18)
    (53,1.322201e-18)
    (54,9.309621e-19)
};
\addlegendentry{fp64 DSA}
\end{axis}
\end{tikzpicture}
    \caption{Fission-source error for SI-only, FP32 DSA, and FP64 DSA calculations of the three-dimensional Jezebel problem. FP32 DSA behaves exactly like FP64 DSA while requiring only half the memory in the DSA solver.}
    \label{fig:jez3d_keff_ardra}
\end{figure}

A 48-group three-dimensional version of Jezebel was considered. The problem had 60 zones in each spatial dimension for a total of 216,000 zones.
The problem was discretized using diamond differencing in space, $S_{8}$ level-symmetric quadrature in angle, and a fourth-order Legendre expansion for the scattering operator. Evaluated Nuclear Data Library (ENDL) 2009 version 5.0 nuclear data were used for this problem. The problem was decomposed across 64 MPI processes (four in each spatial dimension).
Figure \ref{fig:jez3d_keff_ardra} shows the fission-source error, defined as the $L_{1}$ norm of the difference between two fission-source iterates, for an SI-only calculation and calculations using FP32 and FP64 DSA. The convergence tolerance was set to $1 \times 10^{-18}$ to determine whether FP32 DSA would stagnate. In this case, FP32 and FP64 DSA were identical and reduced the number of outer iterations by approximately 45.
Memory edits in ARDRA indicated that FP32 DSA required 292.69~megabytes of memory during one outer iteration, compared with 585.38~megabytes for FP64 DSA. These values are not high-water marks because memory may be freed as different energy groups are completed.

\subsubsection{Two-Dimensional Axisymmetric ($r$--$z$) Jezebel}
An axisymmetric cylindrical ($r$--$z$) variant of Jezebel was also considered. The problem had 160 zones in $r$ and 320 zones in $z$, for a total of 51,200 zones.
The problem was discretized using diamond differencing in space, block Jacobi in energy, and a product quadrature in angle with 12 Gaussian polar levels and 16 uniformly spaced azimuthal angles per level.
The problem was decomposed over 16 MPI processes (four in each spatial dimension) and used ENDL 2009 version 5.0 nuclear data. The convergence tolerance was set to $1 \times 10^{-16}$.
For this problem, DSA was ineffective as an accelerator, but the FP32 and FP64 DSA solvers performed identically. Memory edits in ARDRA indicated that FP32 DSA required 42.61~megabytes, compared with 85.22~megabytes for FP64 DSA.

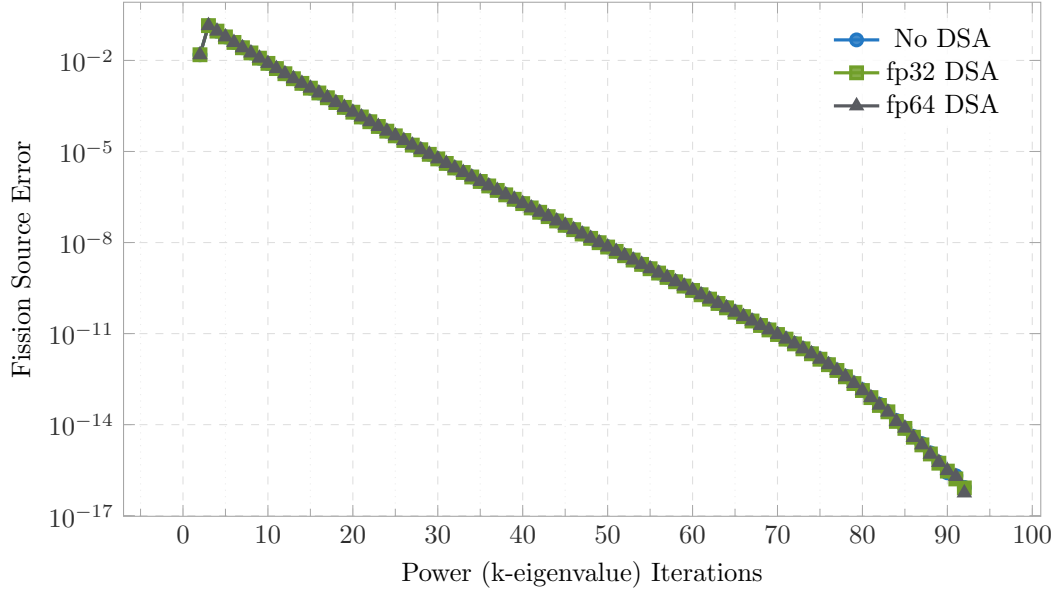
\begin{figure}
    \centering
    \definecolor{llnlblue}{HTML}{2378C3}
\definecolor{llnlgreen}{HTML}{719F2A}
\definecolor{llnlgray}{HTML}{595B60}
\definecolor{llnlorange}{HTML}{D55E00}
\definecolor{llnldarkblue}{HTML}{005DAB}
\definecolor{llnlpurple}{HTML}{CC79A7}
\definecolor{llnllightblue}{HTML}{56B4E9}
\definecolor{llnlblack}{HTML}{000000}

\begin{tikzpicture}
\begin{axis}[
    width=0.90\linewidth,
    height=0.55\linewidth,
    xlabel={Power (k-eigenvalue) Iterations},
    ylabel={Fission Source Error},
    title style={color=llnldarkblue, font=\bfseries},
    axis line style={draw=gray!70},
    tick style={draw=gray!70},
    tick label style={color=gray!35!black},
    legend pos=north east,
    legend style={draw=none, fill=white, fill opacity=0.85, text opacity=1},
    grid=major,
    major grid style={draw=gray!25, dashed},
    minor grid style={draw=gray!15, dotted},
    minor x tick num=1,
    ymode=log,
    log basis y=10,
    minor y tick num=8,
    grid=both,
    ymin=9.5503904752e-18,
    ymax=0.826717340743
]

\addplot[color=llnlblue, line width=1.2pt, mark=*, mark size=2.2pt]
coordinates {
    (2,0.01538828)
    (3,0.1404359)
    (4,0.09242763)
    (5,0.05966028)
    (6,0.03916934)
    (7,0.0262251)
    (8,0.01769444)
    (9,0.01190199)
    (10,0.008017718)
    (11,0.005410698)
    (12,0.003659777)
    (13,0.002508559)
    (14,0.001761896)
    (15,0.001224862)
    (16,0.000850981)
    (17,0.0005921061)
    (18,0.0004118271)
    (19,0.0002855289)
    (20,0.0001983054)
    (21,0.0001380116)
    (22,9.624379e-05)
    (23,6.725015e-05)
    (24,4.708339e-05)
    (25,3.302726e-05)
    (26,2.321019e-05)
    (27,1.634005e-05)
    (28,1.152288e-05)
    (29,8.138868e-06)
    (30,5.757356e-06)
    (31,4.078489e-06)
    (32,2.893035e-06)
    (33,2.054698e-06)
    (34,1.460983e-06)
    (35,1.039943e-06)
    (36,7.409808e-07)
    (37,5.284522e-07)
    (38,3.772031e-07)
    (39,2.695923e-07)
    (40,1.941802e-07)
    (41,1.40035e-07)
    (42,1.006922e-07)
    (43,7.241227e-08)
    (44,5.208239e-08)
    (45,3.746489e-08)
    (46,2.695304e-08)
    (47,1.939253e-08)
    (48,1.395408e-08)
    (49,1.004168e-08)
    (50,7.226726e-09)
    (51,5.201389e-09)
    (52,3.743809e-09)
    (53,2.694749e-09)
    (54,1.939581e-09)
    (55,1.396074e-09)
    (56,1.00494e-09)
    (57,7.233498e-10)
    (58,5.207062e-10)
    (59,3.748075e-10)
    (60,2.697579e-10)
    (61,1.940703e-10)
    (62,1.396525e-10)
    (63,1.003982e-10)
    (64,7.214976e-11)
    (65,5.179819e-11)
    (66,3.713055e-11)
    (67,2.662116e-11)
    (68,1.893664e-11)
    (69,1.3529e-11)
    (70,9.569513e-12)
    (71,6.719285e-12)
    (72,4.739884e-12)
    (73,3.217322e-12)
    (74,2.226328e-12)
    (75,1.475171e-12)
    (76,9.779292e-13)
    (77,6.319099e-13)
    (78,3.950834e-13)
    (79,2.339948e-13)
    (80,1.409918e-13)
    (81,7.862944e-14)
    (82,4.688011e-14)
    (83,2.68144e-14)
    (84,1.368867e-14)
    (85,7.74777e-15)
    (86,4.080485e-15)
    (87,2.365209e-15)
    (88,1.170027e-15)
    (89,5.746693e-16)
    (90,2.593831e-16)
    (91,2.069494e-16)
    (92,8.561465e-17)
};
\addlegendentry{No DSA}

\addplot[color=llnlgreen, line width=1.2pt, mark=square*, mark size=2.2pt]
coordinates {
    (2,0.01539941)
    (3,0.1404064)
    (4,0.092379)
    (5,0.05960565)
    (6,0.03911974)
    (7,0.02618346)
    (8,0.01765977)
    (9,0.01187462)
    (10,0.007996575)
    (11,0.005394653)
    (12,0.0036477)
    (13,0.00249991)
    (14,0.001755197)
    (15,0.001219679)
    (16,0.000847065)
    (17,0.0005891604)
    (18,0.0004096363)
    (19,0.0002839145)
    (20,0.0001971168)
    (21,0.0001371376)
    (22,9.560135e-05)
    (23,6.677829e-05)
    (24,4.673693e-05)
    (25,3.277291e-05)
    (26,2.302348e-05)
    (27,1.6203e-05)
    (28,1.142226e-05)
    (29,8.064988e-06)
    (30,5.703103e-06)
    (31,4.038643e-06)
    (32,2.863766e-06)
    (33,2.033195e-06)
    (34,1.445184e-06)
    (35,1.028333e-06)
    (36,7.324496e-07)
    (37,5.221824e-07)
    (38,3.725953e-07)
    (39,2.66212e-07)
    (40,1.916834e-07)
    (41,1.381777e-07)
    (42,9.93195e-08)
    (43,7.139863e-08)
    (44,5.13342e-08)
    (45,3.691297e-08)
    (46,2.654612e-08)
    (47,1.909264e-08)
    (48,1.373322e-08)
    (49,9.879052e-09)
    (50,7.107165e-09)
    (51,5.113303e-09)
    (52,3.679093e-09)
    (53,2.647245e-09)
    (54,1.904625e-09)
    (55,1.370455e-09)
    (56,9.860973e-10)
    (57,7.095896e-10)
    (58,5.105281e-10)
    (59,3.673894e-10)
    (60,2.64287e-10)
    (61,1.901146e-10)
    (62,1.366939e-10)
    (63,9.827005e-11)
    (64,7.05876e-11)
    (65,5.064707e-11)
    (66,3.627232e-11)
    (67,2.598733e-11)
    (68,1.849741e-11)
    (69,1.320192e-11)
    (70,9.326476e-12)
    (71,6.572496e-12)
    (72,4.578291e-12)
    (73,3.14236e-12)
    (74,2.161941e-12)
    (75,1.431131e-12)
    (76,9.448393e-13)
    (77,6.094669e-13)
    (78,3.767524e-13)
    (79,2.257254e-13)
    (80,1.331198e-13)
    (81,7.774186e-14)
    (82,4.284968e-14)
    (83,2.644592e-14)
    (84,1.279179e-14)
    (85,7.530102e-15)
    (86,3.84766e-15)
    (87,2.152861e-15)
    (88,1.082265e-15)
    (89,5.368621e-16)
    (90,2.94509e-16)
    (91,1.625243e-16)
    (92,8.220788e-17)
};
\addlegendentry{fp32 DSA}

\addplot[color=llnlgray, line width=1.2pt, mark=triangle*, mark size=2.2pt]
coordinates {
    (2,0.01539941)
    (3,0.1404064)
    (4,0.092379)
    (5,0.05960565)
    (6,0.03911974)
    (7,0.02618346)
    (8,0.01765977)
    (9,0.01187462)
    (10,0.007996575)
    (11,0.005394653)
    (12,0.0036477)
    (13,0.00249991)
    (14,0.001755197)
    (15,0.001219679)
    (16,0.000847065)
    (17,0.0005891604)
    (18,0.0004096363)
    (19,0.0002839145)
    (20,0.0001971168)
    (21,0.0001371376)
    (22,9.560135e-05)
    (23,6.677829e-05)
    (24,4.673693e-05)
    (25,3.277291e-05)
    (26,2.302348e-05)
    (27,1.6203e-05)
    (28,1.142226e-05)
    (29,8.064988e-06)
    (30,5.703103e-06)
    (31,4.038643e-06)
    (32,2.863766e-06)
    (33,2.033195e-06)
    (34,1.445184e-06)
    (35,1.028333e-06)
    (36,7.324496e-07)
    (37,5.221824e-07)
    (38,3.725952e-07)
    (39,2.66212e-07)
    (40,1.916834e-07)
    (41,1.381777e-07)
    (42,9.931952e-08)
    (43,7.139862e-08)
    (44,5.13342e-08)
    (45,3.691296e-08)
    (46,2.654612e-08)
    (47,1.909264e-08)
    (48,1.373322e-08)
    (49,9.879045e-09)
    (50,7.107167e-09)
    (51,5.113303e-09)
    (52,3.67909e-09)
    (53,2.647248e-09)
    (54,1.904626e-09)
    (55,1.37045e-09)
    (56,9.861031e-10)
    (57,7.095834e-10)
    (58,5.105304e-10)
    (59,3.673913e-10)
    (60,2.642865e-10)
    (61,1.901137e-10)
    (62,1.36693e-10)
    (63,9.827342e-11)
    (64,7.058553e-11)
    (65,5.064486e-11)
    (66,3.62725e-11)
    (67,2.598997e-11)
    (68,1.849451e-11)
    (69,1.320791e-11)
    (70,9.320281e-12)
    (71,6.575538e-12)
    (72,4.578807e-12)
    (73,3.140813e-12)
    (74,2.161022e-12)
    (75,1.433005e-12)
    (76,9.443261e-13)
    (77,6.094749e-13)
    (78,3.767566e-13)
    (79,2.26532e-13)
    (80,1.320149e-13)
    (81,7.747884e-14)
    (82,4.385275e-14)
    (83,2.591736e-14)
    (84,1.261501e-14)
    (85,7.683507e-15)
    (86,3.792599e-15)
    (87,2.154414e-15)
    (88,1.058931e-15)
    (89,5.710829e-16)
    (90,3.067314e-16)
    (91,1.959783e-16)
    (92,5.622119e-17)
};
\addlegendentry{fp64 DSA}
\end{axis}
\end{tikzpicture}
    \caption{Fission-source error for SI-only, FP32 DSA, and FP64 DSA calculations of the two-dimensional axisymmetric Jezebel problem. FP32 DSA behaves exactly like FP64 DSA while requiring only half the memory in the DSA solver. Neither DSA option accelerates the problem.}
    \label{fig:jez2drz_keff_ardra}
\end{figure}

\subsubsection{The Three-Dimensional C5G7 Mixed-Oxide Benchmark}
\begin{figure}
    \centering
    \input{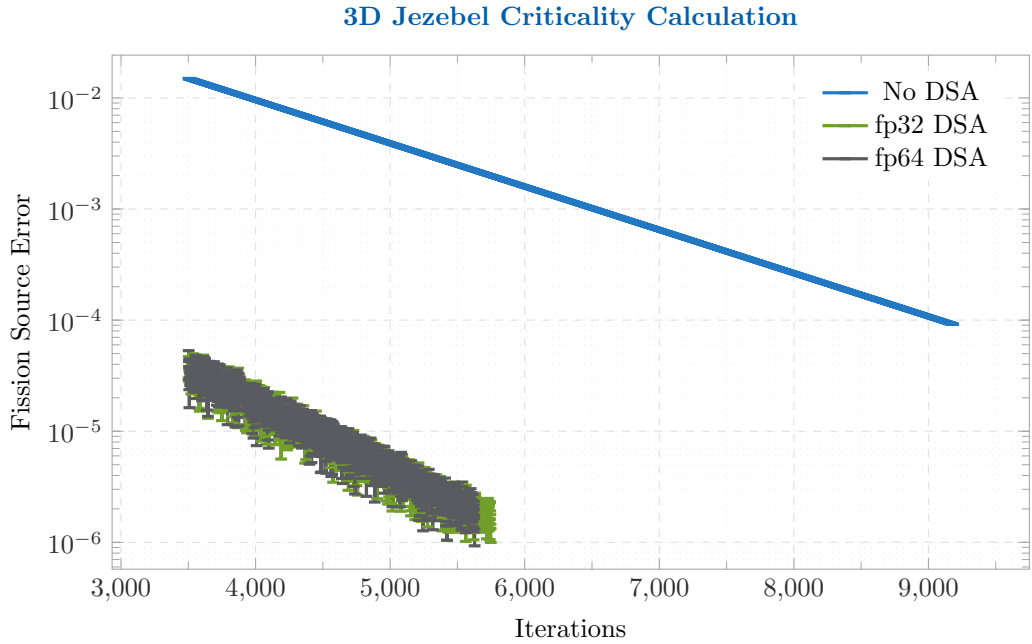}
    \caption{Fission source error for the three-dimensional C5G7 MOX benchmark. FP32 and FP64 DSA exhibit very similar error behavior including similar oscillations. However, both variants of DSA converge in a similar amount of outer iterations.}
    \label{fig:c5g7mox3d_keff_ardra}
\end{figure}

The 3D C5G7 MOX benchmark problem \cite{nea2003c5g7} was simulated using floating-point precision DSA. The ARDRA calculation used $S_{16}$ and had seven energy groups. Both DSA solvers show the similar fission source convergence behavior as shown in Figure \ref{fig:c5g7mox3d_keff_ardra} while accelerating the problem similarly. In this particular problem, ARDRA memory edits show that the cost of FP32 DSA is 4.2 GB of memory while FP64 DSA requires about 8.4 GB. For the three-dimensional problem, ARDRA FP32 requires 5748 outer iterations compared to 5628 outer iterations for FP64 DSA. Timing edits in ARDRA indicate that FP32 DSA wais called 40236 times (seven DSA solves per each outer iteration for 5748 outer iterations) for a total time in the DSA solver of approximately 344 seconds. FP64 DSA was called 39396 times for a total time of 488 seconds. This suggests that FP32 DSA is faster on a per-function-call basis. However, in this problem, the additional overhead of more outer iterations increased total time to solution by approximately 200 seconds. Additional study is required to find the set of conditions where FP32 speed overcomes the cost of additional iterations.

\section{Discussions, Conclusions, \& Future Work}

These results show significant promise for reduced precision DSA.
Across the problems studied, reducing the precision of the DSA solve generally did not substantially degrade iterative performance, and several criticality calculations exhibited nearly identical FP32 and FP64 convergence histories.
The exceptions, including the slower NIF calculation and non-finite FP32 SMG cases, show that robustness still depends on the conditioning and accuracy of the low-order solve.
We therefore conjecture that FP32 DSA is a practical option for many production calculations, provided that poorly conditioned systems are detected and can fall back to FP64.

The measured DSA allocations in the Jezebel problems were reduced by approximately one half.
We conjecture that this reduction will translate into substantial application-level memory savings for problems in which DSA represents a significant fraction of the total memory footprint.
Establishing high-water-mark savings, solver time, and scaling on larger problems is an important subject of future work.
Criticality calculations also require closer study of inexact convergence within each transport sweep before the PBJ iteration couples the energy groups.
An insufficiently converged group sweep can pass error into the between-group iteration and subsequently into the fission source and eigenvalue updates, potentially obscuring or amplifying the effect of FP32 DSA.
Future comparisons should therefore vary the inner sweep tolerance independently of the PBJ and criticality tolerances to determine how accurately each group must be solved for stable and efficient mixed-precision convergence.

\section*{Acknowledgements}
Authors MO and JM thank the high performance computing staff at Lawrence Livermore National Laboratory with support on the rzAdams and rzHound machines as well as the Hypre development team for support with mixed precision tools.
Author JM thanks Peter Brown for useful discussions about ARDRA.

This work performed under the auspices of the U.S. Department of Energy by Lawrence Livermore National Laboratory under Contract DE-AC52-07NA27344.

\bibliographystyle{IEEEtran}
\bibliography{main}

\end{document}